\documentclass[journal]{IEEEtran}

\IEEEoverridecommandlockouts
\usepackage{cite}
\usepackage{amsmath,amssymb,amsfonts}
\usepackage{algorithmic}
\usepackage{graphicx}
\usepackage{textcomp}

\def\BibTeX{{\rm B\kern-.05em{\sc i\kern-.025em b}\kern-.08em
		T\kern-.1667em\lower.7ex\hbox{E}\kern-.125emX}}
\graphicspath{{./mats_images/}}

\begin{document}
\title{Deep Unfolded Recovery of Sub-Nyquist Sampled Ultrasound Images}

\author{\IEEEauthorblockN{Alon Mamistvalov and Yonina C. Eldar}
\thanks{A. Mamistvalov and Y. C. Eldar are with the Faculty of Math and CS,	Weizmann Institute of Science, Rehovot, Israel, email: \{alon.mamistvalov, yonina.eldar\} @weizmann.ac.il.}
\thanks{The work was supported  in part by the Igel Manya Center for Biomedical	Engineering and Signal Processing, as well as the Benoziyo Endowment Fund for the Advancement of Science, the Estate of Olga Klein – Astrachan, and the European Union’s Horizon 2020 research and innovation program under grant No. 646804-ERC-COG-BNYQ;}}



\maketitle

\begin{abstract}
	The most common technique for generating B-mode ultrasound (US) images is delay and sum (DAS) beamforming, where the signals received at the transducer array are sampled before an appropriate delay is applied. This necessitates sampling rates exceeding the Nyquist rate and the use of a large number of antenna elements to ensure sufficient image quality. Recently we proposed methods to reduce the sampling rate and the array size relying on image recovery using iterative algorithms, based on compressed sensing (CS) and the finite rate of innovation (FRI) frameworks. Iterative algorithms typically require a large number of iterations, making them difficult to use in real-time. Here, we propose a reconstruction method from sub-Nyquist samples in the time and spatial domain, that is based on unfolding the ISTA algorithm, resulting in an efficient and interpretable deep network. The inputs to our network are the subsampled beamformed signals after summation and delay in the frequency domain, requiring only a subset of the US signal to be stored for recovery. Our method allows reducing the number of array elements, sampling rate, and computational time while ensuring high quality imaging performance. Using \emph{in vivo} data we demonstrate that the proposed method yields high-quality images while reducing the data volume traditionally used up to 36 times. In terms of image resolution and contrast, our technique outperforms previously suggested methods as well as DAS and minimum-variance (MV) beamforming, paving the way to real-time applicable recovery methods.
\end{abstract}


\section{Introduction}
\IEEEPARstart{M}{edical} ultrasound (US) imaging has become the method of choice in a variety of clinical cases, due to its non-radiating and non-invasive nature while being of low cost compared to other medical imaging modalities. In classic US imaging, an output image is built line by line, by transmitting from an array of transducer elements a series of acoustic pulses along a narrow beam. While propagating through the tissue, the pulses are scattered after encountering acoustic impedance perturbations, resulting in echoes that are detected by the same transducer array. The collected data is then processed digitally resulting in an image line, in a method referred to as beamforming \cite{van1988beamforming}. The beamforming process involves applying an appropriate time delay to the received signals and subsequently averaging them, resulting in a B-mode US image.

Having a relatively low computational load, the most widely used method for beamforming is delay and sum (DAS) ~\cite{thomenius1996evolution,steinberg1992digital}. In DAS the delays are implemented digitally after the data is sampled. However, to avoid artifacts and to obtain fine time delay resolution, US signals are traditionally sampled at a rate 4 to 10 times higher than their actual Nyquist rate \cite{eldar2015sampling}, which is the rate required for perfect reconstruction of the beamformed US signal \cite{shannon1949communication}. These high rates result in a huge amount of data that needs to be stored and processed.

Generating high-resolution US images typically requires that the beam-pattern of the beamformer will consist of a narrow main lobe and low sidelobes \cite{ranganathan2003novel}. To this end, adaptive beamforming methods, such as minimum variance (MV), can be used \cite{capon1969high}. Although MV offers better image resolution, its computational cost is relatively high and increases the data volume. Therefore in clinical scenarios, where real-time imaging is required, the method of choice is DAS beamforming, leading to decreased image quality. Another possible approach for achieving high-resolution US images is considering larger transducer arrays, with more receiving elements, while keeping the pitch below half the wavelength to avoid grating sidelobes \cite{lockwood1998real}. However, a large set of receiving channels, each consisting of high rate samplers, leads to an enormous amount of data to store and process. Therefore, it is of great importance to consider an US imaging method that will generate high-quality images out of partial data, sampled at a low rate.

To reduce the data size, easing the high computational load, but yet reconstructing high quality US images, several techniques have been studied recently. The authors in ~\cite{wagner2012compressed,chernyakova2014fourier} consider the translation of the process of time alignment to the frequency domain. They showed that DAS beamforming can be implemented equivalently in the Fourier domain, leading to a method called frequency domain beamforming (FDBF). Based on FDBF they considered sub-Nyquist sampling while reconstructing the signals using compressed sensing (CS) techniques \cite{eldar2012compressed} based on the finite rate of innovation (FRI) structure of the beamformed signal \cite{tur2011innovation}. 

Other approaches for reducing the number of receiving elements were studied in the literature. These methods consider the usage of sparse arrays ~\cite{roux2017validation,yen2000sparse,austeng2002sparse,brunke1997broad} where some of the elements are removed. In \cite{cohen2018sparse}, a technique for reducing the data size using a sparse array with a new beamforming method called convolutional beamforming algorithm (COBA) is presented. Cohen \emph{et al.} showed that by choosing an array that satisfies appropriate conditions, the achieved beampattern of the virtual array, based on the array's sum co-array ~\cite{cohen2018optimized,cohen2020sparse,liu2017maximally}, is equivalent to the full uniform linear array (ULA). Thus, using a subset of array elements, whose sum co-array is the full ULA, results in better resolution and contrast using fewer receiving elements.

In ~\cite{mamistvalov2020compressed,mamistvalov2020sparse}, we combined the ideas of spatial and sampling rate reduction. High-quality US images were produced by sampling the signals below their Nyquist rate, using a sparse set of array elements, in a method called compressed frequency domain COBA (CFCOBA). For reconstructing the signals, the FRI structure of the convolutionally beamformed signal was exploited \cite{mamistvalov2020compressed}, leading to recovery of the signal based on solving an optimization problem using CS techniques and iterative algorithms.

Many other works, inspired by the impressive performance of deep learning \cite{lecun2015deep}, in a variety of fields including medical imaging \cite{greenspan2016guest}, combined the ideas of deep learning with US imaging \cite{van2019deep}. To the best of our knowledge, all previously suggested deep learning-based methods in US imaging use and store full channel data of all transmitting elements, leading to an enormous size of information to store. In addition, general deep networks for US imaging are traditionally more complex as they need to process high-dimensional tensors. In \cite{zhuang2019deep} for example, the authors combined MV beamforming with deep learning for contrast improvement. Their idea is to use an ensemble of networks, operating over sub-bands in the frequency domain, before or after computing the adaptive weights. The authors in \cite{luijten2020adaptive} considered applying adaptive weights which are calculated using a deep network, with a fully connected deep neural network. In \cite{kessler2020deep}, the authors used temporally and spatially sub-sampled data for generating high-quality US images using a deep neural network that is based on a UNet architecture \cite{ronneberger2015u}. However, they reached only an 11-fold reduction in data size. Moreover, their recovery method required the storage of the diluted data from all of the channels, while in practice, as we will show, only a subset of the Fourier coefficients of the beamformed signal itself, after the summation over the elements, is needed, leading to further reduction of the data size.

Our main goal in this work is to present an efficient recovery method of a beamformed signal from a small subset of its samples, based on a deep unfolded neural network. The proposed method, as opposed to other complicated deep learning approaches for beamforming, consists of fast and efficient training and testing steps since only a small fraction of the samples of the beamformed signal needs to be stored for training the network and for full image recovery. Namely, while previous works needed data from all receiving channels we suggest using subsamples of the beamformed signal. Furthermore, the suggested technique for recovery is based on a relatively small pre-trained deep network, leading to much faster reconstruction when compared to iterative approaches.

We begin by introducing the model of the beamformed signal which follows an FRI structure, and formulate a recovery problem from the sub-sampled beamformed signal. In previous works, the problem was solved using CS techniques with iterative NESTA \cite{nesterov2005smooth,becker2011nesta}. Here, we present a deep network architecture, based on learned ISTA (LISTA) \cite{monga2019algorithm,gregor2010learning}, which unrolls the iterative ISTA algorithm \cite{beck2009fast}. The unrolling approach was shown to be successful with super-resolution in US \cite{van2019deep}. The basic architecture of our network is the same, namely, cascading several iterations of ISTA, resulting in a deep network. However, we slightly modified the network, by adding a layer for final recovery. In addition, we changed the training process to better fit the characteristics of the US signal, by choosing an appropriate loss function that is different than standard mean-squared-error (MSE), traditionally used for LISTA. In our method, channel data is acquired using a sparse set of array elements each sampled below its effective Nyquist rate. The input to our network is the subsamples of the beamformed signal after delay is applied in the frequency domain following \cite{chernyakova2014fourier}. The signals are then summed according to FDBF or CFCOBA, resulting in the input to our network. We train our network with DAS beamformed targets, where training is performed for each image line separately. Despite the data size reduction that reaches 36-times fewer data and the simplification of the training process compared to other deep learning-based approaches, our network consists of only 30 layers that are only half the iterations needed for reconstruction in an iterative manner using NESTA which incorporates more complex steps. Specifically, we present recovery in computation time of $O(n)$ achieving a reduction of the order of square root the operations needed for recovery in NESTA, $O(n^2)$. Moreover, our method shows higher robustness to data size shrinkage by yielding better results in cases of higher data size reduction. Thus, our approach offers much faster and higher quality imaging than those of the existing methods.

The rest of the paper is organized as follows. Section \ref{sec: USImageRec} discusses several previously proposed beamforming methods. Section \ref{sec:model_based} reviews methods for beamformed signal recovery from sub-sampled data based on iterative algorithms and algorithm unrolling. In Section \ref{sec: proposed_method} we present our method and discuss its properties. The performance of the suggested technique is validated in Section \ref{sec: results} using phantom and \emph{in vivo} scans. The paper is concluded in Section \ref{sec: conclusion}.

\section{Ultrasound Beamforming Techniques}
\label{sec: USImageRec}
Using multiple transducer elements for transmission and reception of acoustic pulses, an US image is built for each direction $\theta$, line by line. In transmission, a pulse is transmitted by each element in the array, where the pulses are delayed and scaled so that their sum creates a directional beam propagating at a certain direction through the tissue. The real-time computational complexity in the transmission is negligible since the relevant parameters per angle are calculated offline and saved in tables. On the other hand, the beamforming in reception is more challenging. A line in the image is formed by dynamically delaying the signals received at each element of the array and then averaging them, which leads to signal-to-noise-ration (SNR) enhancement. Here, we describe both the standard beamforming method and techniques that allow a reduction in the number of samples required for imaging.

\subsection{Time Domain Beamforming}
\label{sec: TDBF}
Consider a uniform linear array $U$ comprised of $M$ transducer elements aligned along the x-axis. Let $m_0$ be the reference element, at the origin, and $\delta_m$ denote the distance between $m_0$ and the $m$th element. At $t=0$, a pulse is transmitted by each transducer element, resulting in a beam propagating at direction $\theta$ through the tissue, with the speed of sound in the tissue, $c$. Reflectors in the tissue cause echoes that are received by all array elements at times that depend on their location. Let $\phi_m(t)$ be the received signal at the $m$th element. 
The process of beamforming involves averaging the signals received by different array elements while applying appropriate time shifts for alignment due to the differences in arrival time. The delay that needs to be applied is defined by the geometry of the imaging setup. The aligned signal, $\hat{\phi}_m(t;\theta)$, results from applying an appropriate delay to $\phi_m(t)$, and aligning it to $m_0$ \cite{wagner2012compressed},\cite{jensen1999linear}:
\begin{align} \label{delayeDsignal_DAS}
&\hat{\phi}_m(t;\theta) = \phi_m(\tau_m(t,\theta)), \nonumber \\
&\tau_m = \dfrac{1}{2}(t+\sqrt{t^2-4(\delta_m/c)t\sin \theta +4(\delta_m/c)^2}).
\end{align}
These signals are then averaged to form the beamformed signal for direction $\theta$:
\begin{equation} \label{DAS_eq}
\Phi_{DAS}(t;\theta)= \dfrac{1}{M} \sum^{M}_{m=1} \hat{\phi}_m(t;\theta).
\end{equation}

DAS beamforming is performed digitally. To achieve delays on the order of nanoseconds, the required sampling rate is on the order of hundreds of megahertz \cite{demuth1977frequency}, a requirement that is not feasible. Therefore, US signals are sampled at lower rates, on the order of tens of megahertz and fine delay resolution is obtained by digital interpolation. However, these lower rates are still much higher than the Nyquist rate of the signal, which is defined by twice its' bandwidth \cite{eldar2015sampling}. In general, a well-known rule of thumb is that the sampling rate of the signal should be 4 to 10 times the transducer central frequency. These sampling rates lead to a huge amount of data that needs to be stored and processed.

\subsection{Frequency Domain Beamforming}
\label{sec: FDBF}
To reduce the required sampling rate, applying the delay in the frequency domain was presented in \cite{chernyakova2014fourier}. In addition, it was shown that using a small number of Fourier coefficients of the received signals, beamforming can be performed efficiently. Let $c[k]$ denote the $k$th Fourier series coefficient of $\Phi(t;\theta)$ and let $\hat{c}_m[k]$ be the $k$th Fourier series coefficient of $\hat{\phi}_m(t;\theta)$. Then,
\begin{equation}\label{FDBF_eq_delay_BF}
c[k] = \dfrac{1}{M} \sum^{M}_{m=1} \hat{c}_m[k],
\end{equation}
where $\hat{c}_m[k]$ is obtained by calculating
\begin{equation}
\hat{c}_m[k] = \dfrac{1}{T}\int_{0}^{T} I_{[0,T_B(\theta))}(t)\hat{\phi}_m(t;\theta)e^{\frac{-2\pi j}{T}kt }dt.
\end{equation}
The indicator function, $I_{[a,b)}$, is  equal to $1$ when $a\leq t < b$. Here $T_B(\theta) < T$ is defined in \cite{wagner2012compressed} and $T$ is the pulse penetration depth, the beam is supported on $\left[ 0, T_B(\theta)\right)$.

Following \cite{wagner2012compressed} and \cite{chernyakova2014fourier}, one can write the Fourier coefficients of the delayed signal as
\begin{equation} \label{eq_dist_fun_Fourier}
\hat{c}_m[k] = \sum^{\infty}_{n = -\infty} c_m[k-n]Q_{k,m;\theta}[n],
\end{equation}
where $c_m[k]$ are the Fourier coefficients of the received signal in the $m$th channel before the delay is applied. The variables $Q_{k,m;\theta}[n]$ are the Fourier coefficients of a distortion function, determined solely by the geometry of the imaging setup and can be computed offline once and stored in memory. The delay in the frequency domain is obtained by using the Fourier coefficients of the distortion function. The decay properties of $\left\lbrace Q_{k,m;\theta}[n]\right\rbrace $ and the fact that most of the energy of this set is centered around the DC component lead to the option of replacing the summation in \eqref{eq_dist_fun_Fourier} with a relatively small finite summation. Thus, assuming that $\left\lbrace Q_{k,m;\theta}[n]\right\rbrace$ decays rapidly for $n < -N_1, \; n > N_2$ where $N_1, N_2 \in \mathbb{N}$, leads to
\begin{equation} \label{eq_dist_fun_Fourier_notFinal_finite}
\hat{c}_m[k] = \sum_{n=-N_1}^{N_2} c_m[k-n]Q_{k,m;\theta}[n].
\end{equation}
Combining \eqref{eq_dist_fun_Fourier_notFinal_finite} and \eqref{FDBF_eq_delay_BF} leads to FDBF:
\begin{equation} \label{eq_dist_fun_Fourier_finite}
c[k] = \dfrac{1}{M} \sum^{M}_{m=1} \sum_{n=-N_1}^{N_2} c_m[k-n]Q_{k,m;\theta}[n].
\end{equation}
With appropriate zero padding and applying an inverse Fourier transform to $\left\lbrace c[k]\right\rbrace $, the time domain beamformed signal is obtained.

The main advantage of FDBF is that only the nonzero Fourier coefficients of the received signal are required. Those coefficients are obtained from sub-Nyquist samples of the signal at each receiving element, leading to significant reduction in the sampling rate. To obtain the sub-Nyquist samples of the signal, we follow the mechanism proposed and discussed in \cite{tur2011innovation}, \cite{baransky2014sub}. The output is sampled at its effective Nyquist rate and the required Fourier coefficients are the Fourier transform of the output, using these coefficients, the beamformed signal is recovered as will be discussed in Section \ref{subsec: CS}.

\subsection{Convolutional Beamforming}
\label{sec: COBA}
Using a sparse set of array elements, following COBA, \cite{cohen2018sparse}, \cite{cohen2020sparse}, can lead to substantial data reduction, while producing images at least as good as those obtained with DAS. As was shown in \cite{cohen2018sparse}, using COBA, one can achieve the effective beampattern as would have been obtained using the sum co-array of the given array of transducers \cite{hoctor1990unifying}. 

The beamformed signal in COBA is
\begin{equation} \label{COBA_def}
\Phi_{COBA}(p;\theta) = \sum^{N-1}_{n = -(N-1)} \sum^{N-1}_{m = -(N-1)} u_n(p;\theta)u_m(p;\theta),
\end{equation}
where
\begin{equation} \label{signal_norm}
u_m(p;\theta) = e^{j\angle\hat{\phi}_m(p;\theta)} \sqrt{|\hat{\phi}_m(p;\theta)|},
\end{equation}
with $\hat{\phi}_m(p,\theta)$ denotes the $p$th sample of the delayed signal at channel $m$, sampled at sampling intervals $T_s = \frac{1}{f_s}$, with $\angle\hat{\phi}_m(p,\theta)$ and $|\hat{\phi}_m(p,\theta)|$ being its phase and the magnitude, respectively. Applying \eqref{signal_norm}, ensures that the product of COBA will have the same order of the received signal. An equivalent form of \eqref{COBA_def} is
\begin{equation} \label{COBA_efficient1}
\Phi_{COBA}(p;\theta) = \sum^{2(N-1)}_{n = -2(N-1)}s_n(p;\theta),
\end{equation}
where
\begin{equation} \label{notation_conv_u_s}
s_n(p;\theta) = \sum_{i,j\in U,i+j=n} u_i(p;\theta)u_j(p;\theta),
\end{equation}
with $n = -2(N-1),...,2(N-1)$. Therefore, the vector $\boldmath{s}$, can be calculated by applying lateral convolution on $u$ with itself, after zero padding it to length  $2|M|-1$, i.e. the entries, $s_n$ can be calculated by
\begin{equation}\label{conv_signal}
s(p;\theta) = u(p;\theta)\underset{s}{*}u(p;\theta),
\end{equation}
where $\underset{s}{*}$ stands for the lateral convolution.

The beampattern of DAS beamforming is given by
\begin{equation} \label{DAS_BP}
H_{DAS}(\theta) = \sum_{n=-(N-1)}^{N-1} \exp\left( {-j\omega_0\frac{\delta_n \sin\theta}{c}}\right) ,
\end{equation}
where $\omega_0$ is the central frequency of the transducer and $\delta$ being the distance between two adjacent array elements such that $\delta_n = n\delta$. Following \cite{cohen2018sparse} and based on \eqref{COBA_def} we see that $H_{COBA} = H_{DAS}H_{DAS}$, and it can be written as a single polynomial
\begin{equation} \label{COBA_final_BP}
H_{COBA}(\theta) =  \sum_{n=-2(N-1)}^{2(N-1)} a_n\exp\left( {-j\omega_0\frac{\delta \sin\theta}{c}n}\right),
\end{equation}
where ${a_n}$ are intrinsic apodization weights given by $\boldsymbol{a} = \mathbb{I}_M \underset{s}{*} \mathbb{I}_M$. Here $\mathbb{I}_M$ is a binary vector whose $m$th entry is 1 if $m \in M$. 

The beampattern given by \eqref{COBA_final_BP} is equivalent to that of a DAS beamformer, using the sum co-array of the original ULA, $U$. Due to the fact that effectively the array used for imaging is bigger, improved imaging performance is obtained. Therefore, \cite{cohen2018sparse} suggested using a sparse array, given by the original array after removing several elements, for generating a desired beamformer beampattern, in a method called sparse COBA (SCOBA). The use of fractal arrays ~\cite{puente1996fractal,werner1999fractal,werner2003overview,feder2013fractals,falconer2004fractal}, was discussed in \cite{cohen2020sparse}. The geometry of such arrays is given by
\begin{align} \label{fractal}
&W_0 = {0}, \nonumber \\
&W_{r+1} = \cup_{n\in \mathbb{G}}(W_r + nL^r), \; r\in \mathbb{N},
\end{align}
 where $r$ is the array order, $\mathbb{G}$ is the \emph{generator} array in fractal terminology, and $\min(\mathbb{G}) = 0$. The factor $L$ is given by $L = 2\max(\mathbb{G})+1$, and is referred to as the translation factor. This choice of geometry, leads to the resulting array $W$ which after applying convolutional beamforming provides an effective beampattern with better image resolution and contrast.

\subsection{Compressed Frequency Domain Convolutional Beamforming}
\label{sec: CFCOBA}
Further reduction of the data volume was discussed in \cite{mamistvalov2020compressed}, where both time and spatial dilution are applied. The method is referred to as CFCOBA, and it combines the ideas of FDBF and SCOBA using a fractal array geometry. In that way, the main advantages of both FDBF and SCOBA are obtained, and the beamformed signal is recovered from only a partial set of samples, acquired by a sparse set of array elements while preserving the same image quality.

Following the steps in \cite{mamistvalov2020compressed}, we let $N_{sN}$ be the number of samples acquired by sub-Nyquist sampling of the received signal \cite{tur2011innovation}. The compressed frequency domain signal is defined by
\begin{equation} \label{COBA_no_noramlization}
\hat{\Phi}(n_s)_{CFCOBA} = \sum_{n \in U} \sum_{m \in U} \hat{\phi}_n(n_s)\hat{\phi}_m(n_s),
\end{equation}
where $n_s=0,...,N_{sN}-1$, $N_{sN} < |\mu|$ are the discrete time samples, and $\hat{\phi}_m(n_s)$, $\hat{\phi}_n(n_s)$ are the delayed signals at channels $m$ and $n$ respectively, with the delay applied in the frequency domain. Based on the derivations in \cite{mamistvalov2020compressed}, by calculating the Fourier series coefficients of both sides of \eqref{COBA_no_noramlization} a relation between the sub-sampled Fourier coefficients of the received signal and the Fourier coefficients of the convolutionally beamformed signal is given by
\begin{align} \label{conv_almost_final}
\hat{c}[k]_{CFCOBA} = &N_{sN}\sum_{l \in U}\sum_{m \in U} \sum_{p+q = k} \hat{c}_m[p]\hat{c}_l[q]   \nonumber \\
=&N_{sN}\sum_{l \in U}\sum_{m \in U} (\hat{c}_m*\hat{c}_l)[k],
\end{align}
where the last equation is obtained by setting $q = k-p$, and $\hat{c}_m[p], \hat{c}_l[q]$, the Fourier coefficients at channels $m, l$, are calculated by \eqref{eq_dist_fun_Fourier_notFinal_finite}. 

To efficiently calculate the relation in \eqref{conv_almost_final}, it was shown in \cite{mamistvalov2020compressed}, that due to the linear operations in the temporal dimension, the relation between the Fourier coefficients of the received signals and those of the convolutionally beamformed signal is given by
\begin{equation} \label{final_CFCOBA}
\hat{c}[k]_{CFCOBA} = N_{sN} \sum_{l \in S_{\tilde{U}}} (\hat{c} \underset{s}{*}* \hat{c})_l[k],
\end{equation}
where $\tilde{U}$ is a sparse array based on a sparse set of elements from the ULA, $U$, that was previously defined, and $S_{\tilde{U}}$ is its corresponding sum co-array \cite{hoctor1990unifying}. The calculation in \eqref{final_CFCOBA} denotes a two-dimensional convolution operation, one over the temporal dimension and one over the spatial dimension. The calculation is made efficiently using IFFT and appropriate zero padding based on the convolution theorem.

In \cite{mamistvalov2020compressed}, the FRI structure of the convolutionally beamformed signal was proven, and it was shown that it consists of time-shifted replicas of the square of the known transmitted pulse. This leads to the possibility of reconstructing the signal based on a portion of its Fourier coefficients. 

In the next section, we discuss a possible iterative recovery approach of the sub-sampled signal in both methods, FDBF and CFCOBA.

\subsection{Compressed Sensing Recovery for Further Reduction}
\label{subsec: CS}
When sampling below the effective Nyquist rate of the signal, applying IFFT is insufficient. In ~\cite{chernyakova2014fourier,mamistvalov2020compressed}, CS techniques were suggested to fully recover the signal, based on its FRI structure. We follow here the steps in \cite{chernyakova2014fourier} but the same steps are performed in \cite{mamistvalov2020compressed}, using the square of the known transmitted pulse.

According to \cite{wagner2012compressed}, a beamformed signal obeys an FRI model, which means that it can be written as a sum of shifted replicas of the known transmitted pulse, $h(t)$, with unknown amplitudes and delays
\begin{equation} \label{BMF_FRI}
\hat{\Phi}(t) = \sum_{s = 1}^{S} a_sg(t-t_{s}).
\end{equation}
Here $S$ is the number of scattering elements in the tissue in a certain direction $\theta$, and $\left\lbrace a_s \right\rbrace, \left\lbrace t_s \right\rbrace$ are the unknown amplitudes of the reflections and the times at which the reflection from the $s$th scatterer arrived at the receiving element, respectively. Based on the FRI model and the known pulse shape, the beamformed signal is completely defined by $2S$ parameters. For a duration $T$, the Fourier series expansion of \eqref{BMF_FRI} can be written as
\begin{align} \label{FRI_Fourier}
\hat{c}[k] &= \frac{1}{T}\int_{0}^{T}\sum_{s = 1}^{S} a_sh(t-t_{s})e^{j\left( (2\pi)/T\right) kt}dt \\ \nonumber
&= H\left( \frac{2\pi k}{T}\right) \sum_{s = 1}^{S}a_se^{-j\left( (2\pi)/T\right) kt_{s}},
\end{align} 
where $\hat{c}[k]$ are the Fourier coefficients of the beamformed signal, and $H(\omega)$ is the continuous time Fourier transform of $h(t)$. Writing \eqref{FRI_Fourier} in matrix form results in  
\begin{equation} \label{eq:opt_claim}
\boldmath{\hat{c}} = \boldsymbol{HVa},
\end{equation}
where $\hat{\boldsymbol{c}}$ is a vector of length $|\mu|$ with $\left\lbrace  \hat{c}[k] \right\rbrace _{k \in \mu} $ as its entries, $\boldsymbol{H}$ is a diagonal matrix of size $|\mu| \times |\mu|$ with $H(\frac{2\pi k}{T})$ on its diagonal, $\boldsymbol{V}$ is a Fourier matrix of size $|\mu| \times S$ with $ (k,s) $th element $e^{-j\left( (2\pi)/T\right) kt_{s}}$, and $\boldsymbol{a}$ is a vector of size $S$ with the amplitudes, $\left\lbrace a_s \right\rbrace $, as its entries. If the set of ${k}$ is comprised consecutive indices, then the matrix $V$ holds a Vandermonde form, and has full column rank as long as $|\mu| > S$ and the time delays $t_s \neq t_{s'},\: \forall s \neq s'$. Moreover, if $|\mu| > 2S$ than this problem can be solved for the unknown delays and amplitudes using methods such as annihilating filter \cite{stoica1997introduction}.

In practice, the recovery problem is solved using CS methods. By quantizing the delays with step $T_s = \frac{1}{f_s}$, such that $t_s = q_sT_s$, where $f_s$ is the sampling rate for traditional DAS beamforming, and letting $N = \lfloor T/T_s \rfloor$, the Fourier coefficients can be written as
\begin{equation} \label{FRI_Fourier_quant}
\hat{c}[k] = H\left( \frac{2\pi k}{T}\right) \sum_{s = 0}^{N-1}\tilde{a}_se^{-j\left( (2\pi)/N\right) ks}.
\end{equation} 
The vector $\boldsymbol{\tilde{a}}$ of length $N$ is defined as
\begin{align} \label{a_quant}
\tilde{a}_s =
\begin{cases}
{a}_s   ,& \text{if} \quad  s=q_s,\\
0   ,& \text{else.} 
\end{cases}
\end{align}
The recovery problem then reduces to determining the $S$-sparse vector $\boldsymbol{\tilde{b}}$ from
\begin{equation} \label{eq:CS_problem}
\boldmath{\hat{c}} = \boldsymbol{HD\tilde{a}} = \boldsymbol{A\tilde{a}},
\end{equation}
where $\boldsymbol{D}$ is a $|\mu| \times N$ matrix, formed by taking the set $\mu$ of rows from an $N \times N$ FFT matrix. By choosing $|\mu| \geq CL(\log N)^4$ rows uniformly at random for some constant $C > 0$, the matrix $\boldsymbol{A}$ obeys the RIP with high probability \cite{rudelson2008sparse}. The proposed formulation has a standard CS structure.

\begin{figure*}[h!]
	\includegraphics[width=\textwidth,height=5cm]{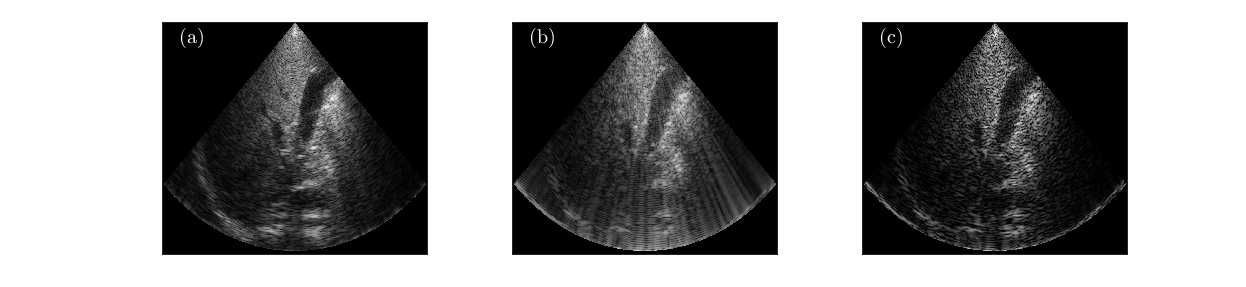}
	\caption{\emph{in-vivo} images produced with: (a) DAS - 1920 samples, (b) FDBF + ISTA recovery - 130 samples, 100 iterations, (c) FDBF + NESTA recovery - 130 samples, 60 iterations.}
	\label{fig:ISTA_res}
\end{figure*}

In practice, due to speckle, the coefficient vector $\boldsymbol{a}$, defined in \eqref{a_quant}, is only approximately sparse. To reconstruct $\boldsymbol{a}$ the $l_1$ norm is used, leading to
\begin{equation} \label{optimization}
\min _{\tilde{a}} ||\boldsymbol{\tilde{a}}||_1 \quad s.t \quad ||\boldsymbol{A\tilde{a} - \hat{c}}||_2 \leq \epsilon,
\end{equation}
with $\epsilon$ being an appropriate noise level.
This optimization problem can be solved using various known techniques, such as interior-point methods \cite{cands20071} or iterative shrinkage techniques \cite{beck2009fast}, \cite{hale2007fixed}. For solving this problem, in \cite{chernyakova2014fourier}, Chernyakova \emph{et al.} suggested to use the NESTA algorithm \cite{nesterov2005smooth}, \cite{becker2011nesta}. This algorithm was shown to be highly suitable for a compressible signal with high dynamic range, which is true in the case of US signals. However, applying NESTA to signals sampled at very low rates leading to a small number of samples fails to fully recover the signal and artifacts, such as high granularity, are present in the resulting image. Moreover, NESTA's computational complexity is relatively high compared to other iterative algorithms, such as ISTA. As will be discussed in Section \ref{sec:ISTA}, ISTA consists of only 2 matrix multiplications, and 1 non-linear operation per iteration; one iteration of NESTA includes more than twice the operations in ISTA, making it less efficient. In addition, in Section \ref{sec: results} we estimate the computational load for NESTA and compare it to the proposed method. For easing the computational load one may suggest using ISTA for recovering the signal, however, as depicted in Fig. \ref{fig:ISTA_res}, even after a relatively high number of 100 iterations, the algorithm fails to recover the beamformed signal, and many artifacts are present in the image.

To overcome the difficulties of recovering the signal from a very small set of samples and the computational load, we next propose a deep unfolded recovery. In Section \ref{sec: results}, we show that for several reduction factors in data size, ranging from 8 to 36 fold dilution, the proposed method performs better than previously suggested iterative algorithms in terms of image quality such as resolution and contrast. Moreover, the suggested technique is more efficient than the iterative techniques, and requires only 30 layers of a deep network for recovering the beamformed signal. Lastly, when comparing our method to traditional end-to-end DNN's, which typically require storing the full channel data prior to delay and sum, in our method we suggest storing only a small set of samples of the beamformed signal, leading to a big reduction in data size.

\section{Model Based Recovery via A Deep Unfolded Network}
\label{sec:model_based}
Iterative algorithms that take into account the structure and the model of the beamformed signal, such as ISTA and NESTA, are widely used in US imaging. These algorithms run iteratively until a certain threshold condition is met. However, they suffer from high computational load as well as artifacts and low image quality when the number of samples is very low. By unfolding the iterations of these algorithms as deep network layers, in a method that is referred to as algorithm unfolding or algorithm unrolling ~\cite{monga2019algorithm,gregor2010learning,van2019deep}, the quality of the recovery is enhanced. In this section, we briefly present the ISTA algorithm for sparse recovery and discuss a possible way of cascading its iteration to a deep network resulting in LISTA. The network proposes the usage of an architecture dedicated to sparse recovery, which is consisted of a relatively small number of layers compared to the number of iterations needed for ISTA and NESTA \cite{van2019deep}. 

\subsection{Iterative Shrinkage Threshold Algorithm}
\label{sec:ISTA}
\begin{figure}[t]
	\includegraphics[width=0.5\textwidth,height=3.6cm]{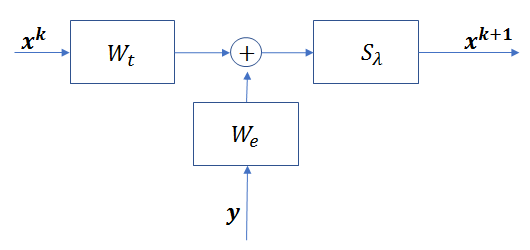}
	\caption{Block diagram of the $k$th iteration of the standard ISTA algorithm.}
	\label{fig:ISTA_block}
\end{figure}

The ISTA algorithm is a widely used algorithm for sparse inference. The convex optimization problem presented in \eqref{optimization}, can be written in an equivalent form as
\begin{equation} \label{optimization_LISTA}
\min_{x} \frac{1}{2}||\boldsymbol{y - Ax}||^2_2 + \lambda||\boldsymbol{x}||_1,
\end{equation}
for a given input vector $\boldsymbol{y} \in \mathbb{C}^{n}$, where $\boldsymbol{A} \in \mathbb{C}^{n\times m}$, with $m \ge n$, is the measurement matrix and $\boldsymbol{x} \in \mathbb{C}^{m}$ is the sparse code representation based on the measurement matrix $\boldsymbol{A}$, and $\lambda$ is a regularization parameter that controls the sparseness of the solution. The block diagram of ISTA is presented in Fig. \ref{fig:ISTA_block}. The main part of ISTA is recursively iterating until convergence, leading to iterations:
\begin{equation}
	\boldsymbol{x}^{k+1} = S_\lambda(\boldsymbol{W_ey + W_tx}^k)), \: \boldsymbol{x}^0 = 0,
\end{equation}
where $\boldsymbol{x}^{k+1}, \boldsymbol{x}^k$ describe the $k$th iteration approximation for $\boldsymbol{x}$, $\boldsymbol{W}_e = \mu \boldsymbol{A}^T$, and $\boldsymbol{W}_t = \boldsymbol{I} -\mu \boldsymbol{W}^T\boldsymbol{W}$ with step size $\mu$, and $S_\lambda$ is a soft thresholding operator with vector of threshold $\lambda$.

\subsection{Learned Iterative Shrinkage Threshold Algorithm}
\label{sec:LISTA}
To overcome the disadvantages of the iterative algorithms, such as high computational load and the difficulty of recovering the signal from a small number of samples, one can unfold the algorithm's iterations into network layers, with each layer replacing the standard algorithm iteration. Unfolding was found to be beneficial in cases when the measurement matrix is not known exactly, which can be the case in US imaging when the scanning depth is changed \cite{monga2019algorithm}, and was also shown to be useful in super-resolution tasks \cite{van2019deep}.

\begin{figure}[t]
	\includegraphics[width=0.5\textwidth,height=3.4cm]{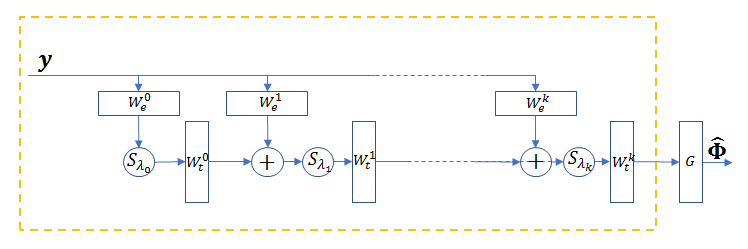}
	\caption{LISTA block diagram. In the proposed network the layer $\boldsymbol{G}$ is added for calculating the output beamformed signal $\boldsymbol{\Phi}$. All rectangles represent $5 \times 1$ 1-D trainable convolution layers, $S_{\lambda_k}$ is a non-linear layer with trainable parameter $\lambda_k$, given by \eqref{eq: proximal} in the proposed architecture.}
	\label{fig:LISTA_block}
\end{figure}

Learned ISTA, LISTA, is based on the ISTA algorithm whose iteration is depicted in Fig. \ref{fig:ISTA_block}. The iteration can be easily converted into a single deep network layer since it comprises a series of analytic operations (summation, matrix-vector multiplication, and soft-thresholding), which are of the same nature as basic operations of a deep neural network. Therefore, executing $K$ iterations of ISTA can be converted to a network of $K$ such cascaded layers. In the unfolded network of LISTA, each layer $k$ of the network consists of trainable convolutions, $\boldsymbol{W}_e^k$, and $\boldsymbol{W}_t^k$ and a trainable shrinkage parameter $\lambda_k$, corresponding to the threshold in ISTA. The block diagram of LISTA is shown in Fig. \ref{fig:LISTA_block}, within the dashed yellow rectangle.

Training is performed with a sequence of input vectors, ${\boldsymbol{y}^1,...,\boldsymbol{y}^N} \in \mathbb{R}^n $ and their corresponding ground-truths sparse codes ${\tilde{\boldsymbol{x}}^1,...,\tilde{\boldsymbol{x}}^N}$. The training process is done by feeding an input, $\boldsymbol{y}^i, \; i \in {1,...,N}$  to the network and retrieving its corresponding predicted sparse code, $\hat{\boldsymbol{x}}^i(\boldsymbol{y}^i;\boldsymbol{W}_t,\boldsymbol{W}_e,\lambda)$. The output is then compared to the ground-truth sparse code, $\tilde{\boldsymbol{x}}^i$, using a loss function
\begin{equation}
\label{eq:loss_LISTA_original}
L(\boldsymbol{W}_t,\boldsymbol{W}_e,\lambda) = \frac{1}{N}\sum_{i=1}^N ||\hat{\boldsymbol{x}}^i(\boldsymbol{y}^i;\boldsymbol{W}_t,\boldsymbol{W}_e,\lambda) - \tilde{\boldsymbol{x}}^i||_2^2.
\end{equation}
Training is performed through loss minimization, using standard methods such as stochastic gradient descent \cite{lecun2012efficient} to learn the weights $\boldsymbol{W}^k_t,\boldsymbol{W}^k_e,\lambda_k$ for each layer. After training is completed, the predicted sparse code is obtained by inserting an input from a dataset different from the training set.

\section{LISTA for Sub-Sampled Ultrasound Image Reconstruction}
\label{sec: proposed_method}
Here we discuss the recovery of the beamformed signal, from a sparse set of it's Fourier coefficients. Since the signal is sampled below the Nyquist rate, it suffers from aliasing in the time domain, after applying inverse Fourier transform. Our goal is to determine the recovered beamformed signal from the sub-sampled signal using a deep network. Our output is a high-quality B-mode US image, consisting of recovered image lines, one for each $\theta$, the propagating direction.

\subsection{Problem Formulation}
In Section \ref{subsec: CS}, we have seen that a beamformed signal obeys an FRI model, therefore its recovery from a subset of Fourier coefficients is reduced to determining a sparse vector from \eqref{eq:CS_problem}. Here we suggest solving \eqref{eq:CS_problem} in the time domain using LISTA. Therefore, the Fourier coefficients are first transformed back to the time domain, leading to
\begin{equation}
\label{eq:time_BMF}
\boldsymbol{\Phi} = \boldsymbol{G\tilde{a}}.
\end{equation}
Here, $\boldsymbol{\Phi} \in \mathbb{R}^{N_{st}}$ is a vector of samples of the beamformed signal, whose Fourier coefficients are the sub-sampled coefficients, $\boldsymbol{G} \in \mathbb{R}^{N_{st} \times N_{st}}$ is a known convolution matrix with time samples of the known transmitted pulse as its entries, and $\boldsymbol{\tilde{a}} \in \mathbb{R} ^{ N_{st}}$ is the $L$-sparse vector such that $N_{st} \gg L$. We insert the vector $\boldsymbol{\Phi}$ to LISTA, as the input $\boldsymbol{y}$ and result in $\hat{\boldsymbol{x}}$ which is the predicted beamformed signal.

\subsection{Network Architecture}

The main goal of our deep unfolded network is to recover the beamformed signal from temporal and spatial sub-sampled data. Although the final output is an image, the training is performed line by line, and the output is the beamformed signal corresponding to each direction in the image.

The basic architecture of our network is similar to that discussed in Section \ref{sec:LISTA}. The input to our network is the 1-D vector of the sub-sampled beamformed signal, $\boldsymbol{\Phi}$. The output of the proposed network is an estimated vector of the beamformed line, of the same length as the input, denoted by $\hat{\boldsymbol{\Phi}}$. To better suit our recovery problem we made two modifications to LISTA. First, a convolution layer is added to the output layer. The new output layer is designed to act as the final conversion between the sparse code, generated by the traditional LISTA, denoted by $\boldsymbol{\tilde{a}}$, to actual replicas of the transmitted pulse, namely it mimics $\boldsymbol{G}$ in \eqref{eq:time_BMF}. The second change is choosing the soft-thresholding layer, $S_{\lambda_k}$, to be of the form \cite{dardikman2020learned}:
\begin{equation}
\label{eq: proximal}
S_{\lambda_k}(x) = \dfrac{x}{1 + e^{-(|x| - \lambda_k)}},
\end{equation}
where $\lambda_k$ is learned and is different in each layer $k \in \{ 0,...,K \}$ of the network. This term was chosen after empirically yielding good results.

The rest of the network consists of 1-D operations designed to replace ISTA, cascaded to overall 30 layers of a deep network. Similarly to LISTA presented in previous sections, the learned parameters are weights of the convolution layers  $\boldsymbol{W}_e^k$, and $\boldsymbol{W}_t^k$, which are of size $5 \times 1$ and a shrinkage parameter $\lambda_k$ as described in \eqref{eq: proximal}. Therefore, the output of the network is $\hat{\boldsymbol{\Phi}}(\boldsymbol{\Phi};\boldsymbol{W}_t,\boldsymbol{W}_e,\lambda)$.

\subsection{Data Sub-Sampling and Pre-Processing}
\label{sec: Data_sub_sample}
We evaluated our method on three different datasets. Prior to training and testing, the data is transferred through a pre-processing pipeline consists of several steps. Two of the datasets consist of temporal sub-sampled data using the Xampling mechanism, \cite{tur2011innovation}, with different reduction factors. The third dataset consists of spatial reduction as well, following ~\cite{cohen2018sparse,cohen2020sparse,mamistvalov2020compressed}. The subsampled Fourier coefficients are calculated as previously described in Section \ref{sec: FDBF}, for the first two datasets, and following Section \ref{sec: CFCOBA} for the third dataset. After calculating the Fourier coefficients, each dataset is transformed back to the time domain, by restoring its negative side spectrum, and zero padding with appropriate zero-vector size, to maintain the same temporal resolution in time as the DAS beamformed signal. Finally, applying inverse Fourier transform results in the sub-sampled signal in the time domain, which is the input to our proposed network, described in the previous section.

\subsection{Loss Function}
In US imaging, the data is characterized by high dynamic range, and is typically being compressed after beamforming for generating the final image. Therefore, we follow \cite{luijten2020adaptive,kessler2020deep}, and replace the MSE function used for LISTA traditionally, \eqref{eq:loss_LISTA_original}, and use signed-mean-squared-logarithmic-error (SMSLE) as the loss function, given by
\begin{align}
\label{eq:Loss_func}
L_{SMSLE}(\boldsymbol{W}_t,&\boldsymbol{W}_e,\lambda) = \\ \nonumber
 &\frac{1}{2}||\log_{10}(\hat{\Phi}^+_{LISTA}) - \log_{10}(\Phi^+_{DAS})|| + \\
			&\frac{1}{2}||\log_{10}(\hat{\Phi}^-_{LISTA}) - \log_{10}(\Phi^-_{DAS})||, \nonumber
\end{align} 
where $\hat{\Phi}^+_{LISTA}, \hat{\Phi}^-_{LISTA}$ are the positive and the negative parts of the network's prediction respectively, and $\hat{\Phi}^+_{DAS}, \hat{\Phi}^-_{DAS}$ are the DAS target's positive and negative parts.


\section{Evaluation Results}
\label{sec: results}
In this section we evaluate our results both quantitatively and visually. For quantitative evaluation of our model we discuss its benefits compared to standard DAS and iterative techniques. For evaluating axial and lateral resolution Full-Width-at-Half-Maxima (FWHM) is calculated over phantom scan and \emph{in-vivo} data. 

For contrast evaluation, contrast-to-noise-ratio (CNR) \cite{rodriguez2019generalized}, is calculated. CNR is evaluated from two regions in each image, the cyst mimicking part and it's background. The value is calculated after envelope detection and log-compression, and is given by
\begin{equation}
CNR = 20\log_{10}\left( \dfrac{|\mu_c - \mu_b|}{\sqrt{\sigma_c^2+\sigma_b^2}} \right).
\end{equation}
Here $\mu_b, \mu_c, \sigma_b, \sigma_c$ denote the means and the STD of the cyst and the background, respectively.

The proposed method was evaluated using \emph{in-vivo} data from healthy volunteers for training and testing. The set for testing included tissue-mimicking phantoms Gammex 403GSLE and 404GSLE, as well as different body parts, such as bladder, kidney, and liver.

The acquisition was performed using the Verasonics Vantage 256 System, using the 64-elements phased array transducer P4-2v. The frequency response of this probe is centered at 2.72 MHz, and a sampling rate of 10.8 MHz was used, leading to 1920 samples per image line. Two temporally sub-sampled datasets were generated from it, using the Xampling mechanism, and one dataset of both temporal and spatial dilution. In the first, 230 samples out of 1920 were used per image line, leading to an 8-fold reduction in data size. In the second dataset, only 130 samples were used, leading to a 15-fold reduction in the volume of the data. The third dataset was generated by diluting both time and space data, based on the fractal array geometry, where a generator array $\mathbb{G} = \left\lbrace 0,1\right\rbrace $ with array order 4 was taken, leading to 15 elements, and 230 samples at each, resulting in a 36-fold total reduction.

Data from two healthy volunteers is used for training, and data from a third volunteer is used for testing. The training data consists of 72 frames of bladders, each consists of 128 image lines of size 1920. The testing is applied to a wide range of organs and not solely to the bladder, to check the robustness of the proposed method. Targets for training were generated using standard DAS beamforming. DAS was chosen over minimum variance (MV) beamforming, \cite{synnevag2007adaptive}, after empirically yielding better results.

The network was implemented using Keras based on Tensorflow backend. The training was done separately to each of the datasets, based on the Adam optimizer, reaching convergence in less than 120 epochs. The initial learning rate was $1e-3$, and the learning rate was set to decrease each time the loss is not improved over few epochs, to make sure the minimum is reached and not skipped due to high step size. The input vector size to the network was $1920\times1$ as described previously. The initialization of the weights was based on the Glorot Uniform initialization \cite{glorot2010understanding}.

\begin{table}[t]
	\centering
	\caption{Resolution Evaluation}
	\label{table: resolution}
	\begin{tabular}{c c || c | c || c | c}
		\multicolumn{2}{c}{Beamforming technique}  & \multicolumn{2}{c}{\emph{in-vivo} data}& \multicolumn{2}{c}{Phantom data}\\
		\hline
		&			 & axial & lateral & axial & lateral\\
		\hline \hline
		DAS  &1920 samples, 64 elements  & 3.66 & 5.01 & 3.2 & 4.5\\
		\hline 
		MV  &1920 samples, 64 elements   & 2.77 & 4.06& 2.64 & 4.2\\
		\hline 
		NESTA  &230 samples, 64 elements & 4.89 & 5.65& 4.82 & 5.56\\
		\hline 
		NESTA  &130 samples, 64 elements & 5.6 & 5.57& 6 & 5.8\\
		\hline 
		NESTA  &230 samples, 15 elements & 4.89 & 5.65& 3.6 & 4.6\\
		\hline 
		LISTA  &230 samples, 64 elements & 3.37 & 4.38& 4.38 & 5.37\\
		\hline 
		LISTA  &130 samples, 64 elements & 8.51 & 7.36& 9.5 & 7.9\\
		\hline 
		LISTA  &230 samples, 15 elements & 1.92 & 2.77& 1.43 & 2.4\\
		\hline
	\end{tabular}
\end{table}

\begin{figure*}[h!]
	\includegraphics[width=\textwidth,height=7cm]{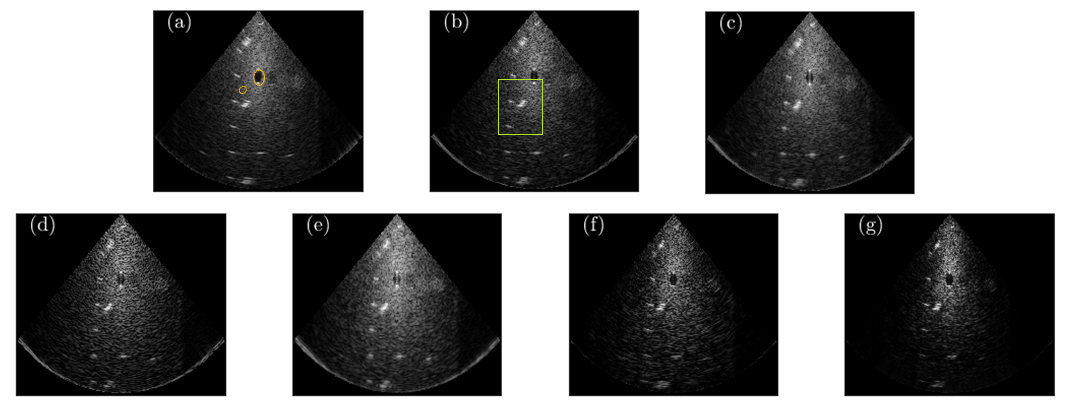}
	\caption{Phantom images produced with: (a) DAS - 1920 samples, (b) FDBF + NESTA recovery - 230 samples, (c) FDBF + LISTA recovery - 230 samples, (d) FDBF + NESTA recovery - 130 samples, (e) FDBF + LISTA recovery - 130 samples, (f) CFCOBA + NESTA recovery - 230 samples, 15 channels, (g) CFCOBA + LISTA recovery - 230 samples, 15 channels. The regions used for calculating CNR are marked with orange circles in (a). The top region denotes the cyst and the bottom the background. In (b) the yellow rectangle marks the zoomed in region in Fig. \ref{fig:pha_res_zoom}.}
	\label{fig:pha_res}
\end{figure*}
\begin{figure*}[h!]
	\centering
	\includegraphics[width=0.9\textwidth,height=5cm]{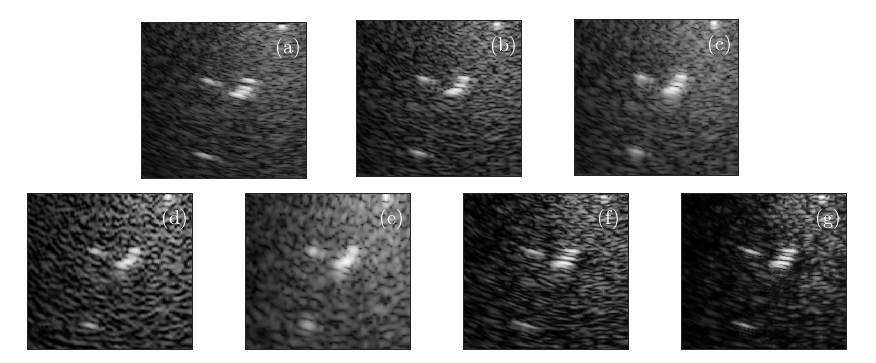}
	\caption{Zoom-in phantom images produced with: (a) DAS - 1920 samples, (b) FDBF + NESTA recovery - 230 samples, (c) FDBF + LISTA recovery - 230 samples, (d) FDBF + NESTA recovery - 130 samples, (e) FDBF + LISTA recovery - 130 samples, (f) CFCOBA + NESTA recovery - 230 samples, 15 channels, (g) CFCOBA + LISTA recovery - 230 samples, 15 channels.}
	\label{fig:pha_res_zoom}
\end{figure*}

\begin{figure*}[h!]
	\includegraphics[width=\textwidth,height=8cm]{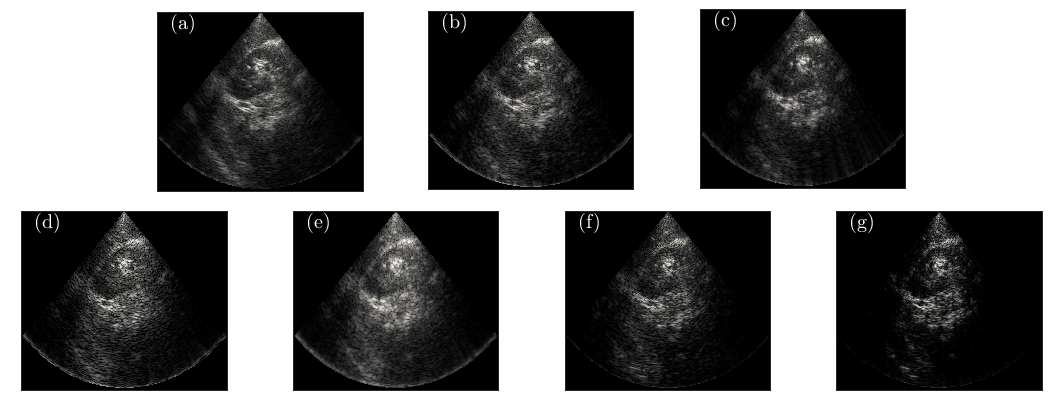}
	\\
	\includegraphics[width=\textwidth,height=8cm]{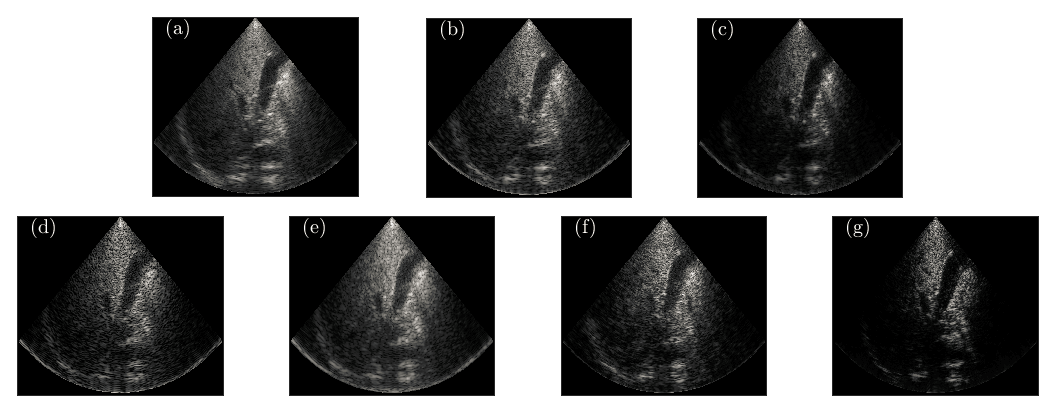}
	\\
	\includegraphics[width=\textwidth,height=8cm]{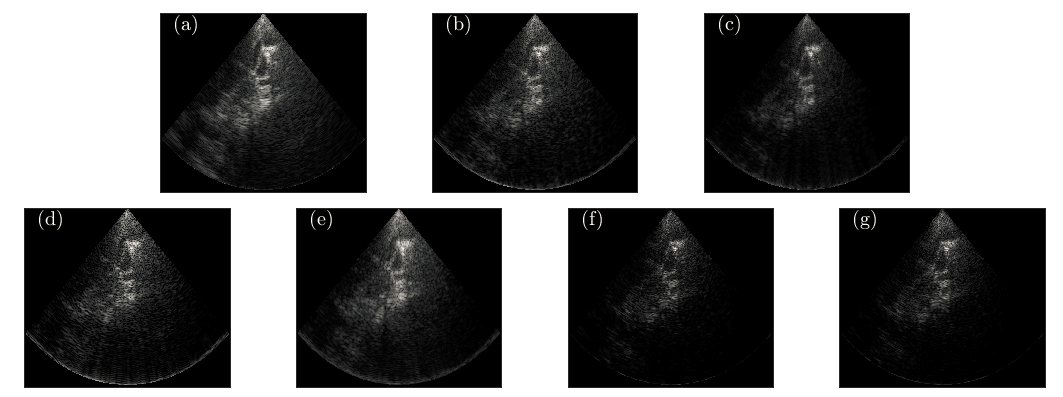}
	\caption{\emph{In-vivo} images produced with: (a) DAS - 1920 samples, (b) FDBF + NESTA recovery - 230 samples, (c) FDBF + LISTA recovery - 230 samples, (d) FDBF + NESTA recovery - 130 samples, (e) FDBF + LISTA recovery - 130 samples, (f) CFCOBA + NESTA recovery - 230 samples, 15 channels, (g) CFCOBA + LISTA recovery - 230 samples, 15 channels.}
	\label{fig:kid_res}
\end{figure*}
To demonstrate the performance of the proposed reconstruction method, we compare it quantitatively to DAS and NESTA, the iterative algorithm used in previous works. 

Table \ref{table: resolution} shows a quantitative evaluation of the lateral and axial resolution, of the compared reconstruction methods. The resolution is calculated from the \emph{in-vivo} scans and averaged over the three organs and for phantom scan data. The results were estimated by computing FWHM for each lateral and axial cut per frame, and respectively averaging the results. For 230 samples at 64 channels, LISTA recovery outperforms NESTA. The recovery for 130 samples using the full ULA, resulted in slightly worse resolution than produced by NESTA, but NESTA caused high granularity in the resulting image. The evaluated resolution of the proposed method over the third dataset is much better than that produced by NESTA, DAS and it is actually better than MV beamforming.

\begin{table}[b]
\centering
\caption{CNR Evaluation}
\label{table: CNR}
\begin{tabular}{c c c }
	\hline
	&			    & CNR (dB)\\
	\hline \hline
	DAS &1920 samples, 64 elements  & 8.35\\
	\hline  
	NESTA &230 samples, 64 elements & 5.9\\
	\hline 
	NESTA &130 samples, 64 elements & 1.9\\
	\hline 
	NESTA &230 samples, 15 elements & 6.1\\
	\hline 
	LISTA &230 samples, 64 elements & 8.7\\
	\hline 
	LISTA &130 samples, 64 elements & 7.7\\
	\hline 
	LISTA &130 samples, 15 elements & 7.9\\
	\hline
\end{tabular}
\end{table}

In Table \ref{table: CNR}, quantitative evaluation of CNR for a tissue-mimicking phantom scan is evaluated. As can be seen, in all cases our method outperforms NESTA. In addition, the second dataset reconstructed using NESTA leads to the lowest CNR, which indicates that contrast between noise the signal of interest almost does not exist. These results indicate that our method is much more robust to a reduction both in sampling rate and in the number of acquiring channels than NESTA reconstruction. The resulting images from the phantom scans are presented in Fig. \ref{fig:pha_res}. For visually better examining the results, zoomed-in images are given in Fig. \ref{fig:pha_res_zoom}. While in (a)-(c) all three options depict approximately the same visual quality. Comparing (d) and (e) should be done more carefully, one may claim that (e) exhibits worse resolution with regards to (d), however, there are two major disadvantages in the latter. First, due to the relatively high reduction in the number of samples per image line, the recovered image shows high granularity, combined with the second drawback, which is low contrast, as calculated and can be seen, earns it difficult to notice actual reflectors that are easily seen in (e). Lastly, in (f)-(g), both contrast and resolution are visually proved to be much better in (g).  

Images produced from \emph{in-vivo} scan data are presented in Fig. \ref{fig:kid_res}. It can be seen that the proposed method, over the first dataset of 230 samples yields results that are visually better than those produced with DAS or NESTA. Moreover, in the case of 130 samples per image line, the proposed method outperforms NESTA. The image reconstructed using NESTA suffers from high granularity that results from the high dilution factor and can be considered as an artifact. In addition, testing the proposed method over the spatially and temporally diluted data presents better image contrast in comparison to NESTA and DAS, a result that agrees with previously calculated CNR.

Finally, in addition to the visual and quantitative improvement of our method, we also managed to ease the computational load compared to the iterative approach used in previous works. While NESTA is an iterative process, our method is based on a pre-trained network that mimics ISTA, a relatively simple algorithm, with weights that are saved before reconstruction, leading to real-time applicable recovery technique since no complicated matrix operations need to be done. Moreover, our method offers a network of only 30 layers, whilst NESTA iterates over 60 times until convergence and produces results not as good when compared to the proposed method. For emphasizing the total easing in computation complexity of our method we look into an estimation of the total operations in each of the methods. The NESTA algorithm includes several matrix multiplications in each iteration, leading to an overall complexity of over $O(n^2)$, for a given input vector of size $n$ and several lower complexity operations. In LISTA each layer includes 2 convolutions with a kernel of size $5 \times 1$ and nonlinear operation, leading to the computational complexity of $O(n)$. This combined with the fact that each iteration of NESTA consists of more operations than the iterative version of ISTA, leads to a highly efficient implementation of US image reconstruction from sub-Nyquist samples of the signal. These results validate that recovery of the beamformed signal using LISTA is done much more efficiently without degrading image quality compared to the iterative approach of NESTA.

\section{Conclusion}
\label{sec: conclusion}
In this paper, we presented a deep unfolding based reconstruction method, for high-quality images, from temporally and spatially sub-sampled channel data based on the Xampling mechanism. We extended the proposed methods in ~\cite{chernyakova2014fourier,mamistvalov2020compressed}, and showed that based on the FRI model of the US signal it can be recovered using a deep network that mimics the iterations of ISTA. 

Our network consists of only 30 folds, leading to an efficient implementation, using a loss function that better fits the studied field of US than traditionally used MSE. After training the network on a dataset of only \emph{in-vivo} bladders, we validated our approach and tested it on phantom channel data and several \emph{in-vivo} scans different from bladders, demonstrating the high versatility of our network. The proposed method produced high quality results both in quantitative and visual terms while using 36 times less data. The output of our technique yields better image resolution and CNR compared to NESTA, proving the robustness of our method, and resulting in visually high-quality B-mode images.

These results prove that our method can be easily plugged into previously suggested schemes, for recovery from sub-Nyquist sampling and sparse arrays, while improving image quality and using an efficient deep neural network instead of iterative algorithms.

\section*{Acknowledgment}
The authors would like to thank Dr. Israel Aharony for voluntarily performing the scans which provided the data for testing and evaluating the proposed method.

\bibliographystyle{ieeetran}
\bibliography{ref}

\begin{thebibliography}{10}
\providecommand{\url}[1]{#1}
\csname url@samestyle\endcsname
\providecommand{\newblock}{\relax}
\providecommand{\bibinfo}[2]{#2}
\providecommand{\BIBentrySTDinterwordspacing}{\spaceskip=0pt\relax}
\providecommand{\BIBentryALTinterwordstretchfactor}{4}
\providecommand{\BIBentryALTinterwordspacing}{\spaceskip=\fontdimen2\font plus
\BIBentryALTinterwordstretchfactor\fontdimen3\font minus
  \fontdimen4\font\relax}
\providecommand{\BIBforeignlanguage}[2]{{%
\expandafter\ifx\csname l@#1\endcsname\relax
\typeout{** WARNING: IEEEtran.bst: No hyphenation pattern has been}%
\typeout{** loaded for the language `#1'. Using the pattern for}%
\typeout{** the default language instead.}%
\else
\language=\csname l@#1\endcsname
\fi
#2}}
\providecommand{\BIBdecl}{\relax}
\BIBdecl

\bibitem{van1988beamforming}
B.~D. Van~Veen and K.~M. Buckley, ``Beamforming: A versatile approach to
  spatial filtering,'' \emph{IEEE assp magazine}, vol.~5, no.~2, pp. 4--24,
  1988.

\bibitem{thomenius1996evolution}
K.~E. Thomenius, ``Evolution of ultrasound beamformers,'' in \emph{1996 IEEE
  Ultrasonics Symposium. Proceedings}, vol.~2.\hskip 1em plus 0.5em minus
  0.4em\relax IEEE, 1996, pp. 1615--1622.

\bibitem{steinberg1992digital}
B.~D. Steinberg, ``Digital beamforming in ultrasound,'' \emph{IEEE Transactions
  on ultrasonics, ferroelectrics, and frequency control}, vol.~39, no.~6, pp.
  716--721, 1992.

\bibitem{eldar2015sampling}
Y.~C. Eldar, \emph{Sampling theory: Beyond bandlimited systems}.\hskip 1em plus
  0.5em minus 0.4em\relax Cambridge University Press, 2015.

\bibitem{shannon1949communication}
C.~E. Shannon, ``Communication in the presence of noise,'' \emph{Proceedings of
  the IRE}, vol.~37, no.~1, pp. 10--21, 1949.

\bibitem{ranganathan2003novel}
K.~Ranganathan and W.~F. Walker, ``A novel beamformer design method for medical
  ultrasound. part i: Theory,'' \emph{IEEE Transactions on ultrasonics,
  ferroelectrics, and frequency control}, vol.~50, no.~1, pp. 15--24, 2003.

\bibitem{capon1969high}
J.~Capon, ``High-resolution frequency-wavenumber spectrum analysis,''
  \emph{Proceedings of the IEEE}, vol.~57, no.~8, pp. 1408--1418, 1969.

\bibitem{lockwood1998real}
G.~R. Lockwood, J.~R. Talman, and S.~S. Brunke, ``Real-time 3-{D} ultrasound
  imaging using sparse synthetic aperture beamforming,'' \emph{IEEE
  transactions on ultrasonics, ferroelectrics, and frequency control}, vol.~45,
  no.~4, pp. 980--988, 1998.

\bibitem{wagner2012compressed}
N.~Wagner, Y.~C. Eldar, and Z.~Friedman, ``Compressed beamforming in ultrasound
  imaging,'' \emph{IEEE Transactions on Signal Processing}, vol.~60, no.~9, pp.
  4643--4657, 2012.

\bibitem{chernyakova2014fourier}
T.~Chernyakova and Y.~C. Eldar, ``Fourier-domain beamforming: the path to
  compressed ultrasound imaging,'' \emph{IEEE transactions on ultrasonics,
  ferroelectrics, and frequency control}, vol.~61, no.~8, pp. 1252--1267, 2014.

\bibitem{eldar2012compressed}
Y.~C. Eldar and G.~Kutyniok, \emph{Compressed sensing: theory and
  applications}.\hskip 1em plus 0.5em minus 0.4em\relax Cambridge university
  press, 2012.

\bibitem{tur2011innovation}
R.~Tur, Y.~C. Eldar, and Z.~Friedman, ``Innovation rate sampling of pulse
  streams with application to ultrasound imaging,'' \emph{IEEE Transactions on
  Signal Processing}, vol.~59, no.~4, pp. 1827--1842, 2011.

\bibitem{roux2017validation}
E.~Roux, E.~Badescu, L.~Petrusca, F.~Varray, A.~Ramalli, C.~Cachard, M.~Robini,
  H.~Liebgott, and P.~Tortoli, ``Validation of optimal 2{D} sparse arrays in
  focused mode: Phantom experiments,'' in \emph{2017 IEEE International
  Ultrasonics Symposium (IUS)}.\hskip 1em plus 0.5em minus 0.4em\relax IEEE,
  2017, pp. 1--4.

\bibitem{yen2000sparse}
J.~T. Yen, J.~P. Steinberg, and S.~W. Smith, ``Sparse 2-{D} array design for
  real time rectilinear volumetric imaging,'' \emph{IEEE transactions on
  ultrasonics, ferroelectrics, and frequency control}, vol.~47, no.~1, pp.
  93--110, 2000.

\bibitem{austeng2002sparse}
A.~Austeng and S.~Holm, ``Sparse 2-{D} arrays for 3-{D} phased array
  imaging-design methods,'' \emph{IEEE transactions on ultrasonics,
  ferroelectrics, and frequency control}, vol.~49, no.~8, pp. 1073--1086, 2002.

\bibitem{brunke1997broad}
S.~S. Brunke and G.~R. Lockwood, ``Broad-bandwidth radiation patterns of sparse
  two-dimensional vernier arrays,'' \emph{IEEE transactions on ultrasonics,
  ferroelectrics, and frequency control}, vol.~44, no.~5, pp. 1101--1109, 1997.

\bibitem{cohen2018sparse}
R.~Cohen and Y.~C. Eldar, ``Sparse convolutional beamforming for ultrasound
  imaging,'' \emph{IEEE transactions on ultrasonics, ferroelectrics, and
  frequency control}, vol.~65, no.~12, pp. 2390--2406, 2018.

\bibitem{cohen2018optimized}
------, ``Optimized sparse array design based on the sum coarray,'' in
  \emph{2018 IEEE International Conference on Acoustics, Speech and Signal
  Processing (ICASSP)}.\hskip 1em plus 0.5em minus 0.4em\relax IEEE, 2018, pp.
  3340--3343.

\bibitem{cohen2020sparse}
------, ``Sparse array design via fractal geometries,'' \emph{arXiv preprint
  arXiv:2001.01217}, 2020.

\bibitem{liu2017maximally}
C.-L. Liu and P.~Vaidyanathan, ``Maximally economic sparse arrays and cantor
  arrays,'' in \emph{2017 IEEE 7th International Workshop on Computational
  Advances in Multi-Sensor Adaptive Processing (CAMSAP)}.\hskip 1em plus 0.5em
  minus 0.4em\relax IEEE, 2017, pp. 1--5.

\bibitem{mamistvalov2020compressed}
A.~Mamistvalov and Y.~C. Eldar, ``Compressed fourier-domain convolutional
  beamforming for wireless ultrasound imaging,'' \emph{arXiv preprint
  arXiv:2010.13171}, 2020.

\bibitem{mamistvalov2020sparse}
------, ``Sparse convolutional beamforming for wireless ultrasound,'' in
  \emph{ICASSP 2020-2020 IEEE International Conference on Acoustics, Speech and
  Signal Processing (ICASSP)}.\hskip 1em plus 0.5em minus 0.4em\relax IEEE,
  2020, pp. 9254--9258.

\bibitem{lecun2015deep}
Y.~LeCun, Y.~Bengio, and G.~Hinton, ``Deep learning,'' \emph{nature}, vol. 521,
  no. 7553, pp. 436--444, 2015.

\bibitem{greenspan2016guest}
H.~Greenspan, B.~Van~Ginneken, and R.~M. Summers, ``Guest editorial deep
  learning in medical imaging: Overview and future promise of an exciting new
  technique,'' \emph{IEEE Transactions on Medical Imaging}, vol.~35, no.~5, pp.
  1153--1159, 2016.

\bibitem{van2019deep}
R.~J. van Sloun, R.~Cohen, and Y.~C. Eldar, ``Deep learning in ultrasound
  imaging,'' \emph{Proceedings of the IEEE}, vol. 108, no.~1, pp. 11--29, 2019.

\bibitem{zhuang2019deep}
R.~Zhuang and J.~Chen, ``Deep learning based minimum variance beamforming for
  ultrasound imaging,'' in \emph{Smart Ultrasound Imaging and Perinatal,
  Preterm and Paediatric Image Analysis}.\hskip 1em plus 0.5em minus
  0.4em\relax Springer, 2019, pp. 83--91.

\bibitem{luijten2020adaptive}
B.~Luijten, R.~Cohen, F.~J. De~Bruijn, H.~A. Schmeitz, M.~Mischi, Y.~C. Eldar,
  and R.~J. Van~Sloun, ``Adaptive ultrasound beamforming using deep learning,''
  \emph{IEEE Transactions on Medical Imaging}, vol.~39, no.~12, pp. 3967--3978,
  2020.

\bibitem{kessler2020deep}
N.~Kessler and Y.~C. Eldar, ``Deep-learning based adaptive ultrasound imaging
  from sub-nyquist channel data,'' \emph{arXiv preprint arXiv:2008.02628},
  2020.

\bibitem{ronneberger2015u}
O.~Ronneberger, P.~Fischer, and T.~Brox, ``U-net: Convolutional networks for
  biomedical image segmentation,'' in \emph{International Conference on Medical
  image computing and computer-assisted intervention}.\hskip 1em plus 0.5em
  minus 0.4em\relax Springer, 2015, pp. 234--241.

\bibitem{nesterov2005smooth}
Y.~Nesterov, ``Smooth minimization of non-smooth functions,''
  \emph{Mathematical programming}, vol. 103, no.~1, pp. 127--152, 2005.

\bibitem{becker2011nesta}
S.~Becker, J.~Bobin, and E.~J. Cand{\`e}s, ``Nesta: A fast and accurate
  first-order method for sparse recovery,'' \emph{SIAM Journal on Imaging
  Sciences}, vol.~4, no.~1, pp. 1--39, 2011.

\bibitem{monga2019algorithm}
V.~Monga, Y.~Li, and Y.~C. Eldar, ``Algorithm unrolling: Interpretable,
  efficient deep learning for signal and image processing,'' \emph{arXiv
  preprint arXiv:1912.10557}, 2019.

\bibitem{gregor2010learning}
K.~Gregor and Y.~LeCun, ``Learning fast approximations of sparse coding,'' in
  \emph{Proceedings of the 27th international conference on international
  conference on machine learning}, 2010, pp. 399--406.

\bibitem{beck2009fast}
A.~Beck and M.~Teboulle, ``A fast iterative shrinkage-thresholding algorithm
  for linear inverse problems,'' \emph{SIAM journal on imaging sciences},
  vol.~2, no.~1, pp. 183--202, 2009.

\bibitem{jensen1999linear}
J.~A. Jensen, ``Linear description of ultrasound imaging systems: Notes for the
  international summer school on advanced ultrasound imaging at the technical
  university of denmark,'' 1999.

\bibitem{demuth1977frequency}
G.~DeMuth, ``Frequency domain beamforming techniques,'' in \emph{ICASSP'77.
  IEEE International Conference on Acoustics, Speech, and Signal Processing},
  vol.~2.\hskip 1em plus 0.5em minus 0.4em\relax IEEE, 1977, pp. 713--715.

\bibitem{baransky2014sub}
E.~Baransky, G.~Itzhak, N.~Wagner, I.~Shmuel, E.~Shoshan, and Y.~C. Eldar,
  ``Sub-{N}yquist radar prototype: Hardware and algorithm,'' \emph{IEEE
  Transactions on Aerospace and Electronic Systems}, vol.~50, no.~2, pp.
  809--822, 2014.

\bibitem{hoctor1990unifying}
R.~T. Hoctor and S.~A. Kassam, ``The unifying role of the coarray in aperture
  synthesis for coherent and incoherent imaging,'' \emph{Proceedings of the
  IEEE}, vol.~78, no.~4, pp. 735--752, 1990.

\bibitem{puente1996fractal}
C.~Puente-Baliarda and R.~Pous, ``Fractal design of multiband and low side-lobe
  arrays,'' \emph{IEEE Transactions on Antennas and Propagation}, vol.~44,
  no.~5, p. 730, 1996.

\bibitem{werner1999fractal}
D.~H. Werner, R.~L. Haupt, and P.~L. Werner, ``Fractal antenna engineering: The
  theory and design of fractal antenna arrays,'' \emph{IEEE Antennas and
  propagation Magazine}, vol.~41, no.~5, pp. 37--58, 1999.

\bibitem{werner2003overview}
D.~H. Werner and S.~Ganguly, ``An overview of fractal antenna engineering
  research,'' \emph{IEEE Antennas and propagation Magazine}, vol.~45, no.~1,
  pp. 38--57, 2003.

\bibitem{feder2013fractals}
J.~Feder, \emph{Fractals}.\hskip 1em plus 0.5em minus 0.4em\relax Springer
  Science \& Business Media, 2013.

\bibitem{falconer2004fractal}
K.~Falconer, \emph{Fractal geometry: mathematical foundations and
  applications}.\hskip 1em plus 0.5em minus 0.4em\relax John Wiley \& Sons,
  2004.

\bibitem{stoica1997introduction}
P.~Stoica, \emph{Introduction to spectral analysis}.\hskip 1em plus 0.5em minus
  0.4em\relax Prentice hall, 2000.

\bibitem{rudelson2008sparse}
M.~Rudelson and R.~Vershynin, ``On sparse reconstruction from fourier and
  gaussian measurements,'' \emph{Communications on Pure and Applied
  Mathematics: A Journal Issued by the Courant Institute of Mathematical
  Sciences}, vol.~61, no.~8, pp. 1025--1045, 2008.

\bibitem{cands20071}
E.~Cands and J.~Romberg, ``l1-magic: Recovery of sparse signals via complex
  programming,'' Technical Report, California Institute of Technology, 2005.,
  Tech. Rep., 2007.

\bibitem{hale2007fixed}
E.~T. Hale, W.~Yin, and Y.~Zhang, ``A fixed-point continuation method for
  l1-regularized minimization with applications to compressed sensing,''
  \emph{CAAM TR07-07, Rice University}, vol.~43, p.~44, 2007.

\bibitem{lecun2012efficient}
Y.~A. LeCun, L.~Bottou, G.~B. Orr, and K.-R. M{\"u}ller, ``Efficient
  backprop,'' in \emph{Neural networks: Tricks of the trade}.\hskip 1em plus
  0.5em minus 0.4em\relax Springer, 2012, pp. 9--48.

\bibitem{dardikman2020learned}
G.~Dardikman-Yoffe and Y.~C. Eldar, ``Learned sparcom: Unfolded deep
  super-resolution microscopy,'' \emph{arXiv preprint arXiv:2004.09270}, 2020.

\bibitem{rodriguez2019generalized}
A.~Rodriguez-Molares, O.~M.~H. Rindal, J.~D’hooge, S.-E. M{\aa}s{\o}y,
  A.~Austeng, M.~A.~L. Bell, and H.~Torp, ``The generalized contrast-to-noise
  ratio: a formal definition for lesion detectability,'' \emph{IEEE
  Transactions on Ultrasonics, Ferroelectrics, and Frequency Control}, vol.~67,
  no.~4, pp. 745--759, 2019.

\bibitem{synnevag2007adaptive}
J.~F. Synnevag, A.~Austeng, and S.~Holm, ``Adaptive beamforming applied to
  medical ultrasound imaging,'' \emph{IEEE transactions on ultrasonics,
  ferroelectrics, and frequency control}, vol.~54, no.~8, pp. 1606--1613, 2007.

\bibitem{glorot2010understanding}
X.~Glorot and Y.~Bengio, ``Understanding the difficulty of training deep
  feedforward neural networks,'' in \emph{Proceedings of the thirteenth
  international conference on artificial intelligence and statistics}, 2010,
  pp. 249--256.

\end{thebibliography}
\end{document}